# Large Scale Integration of Graphene Transistors for Potential Applications in the Back End of the Line

A.D. Smith[1], S. Vaziri[1], S. Rodriguez[1], M. Östling[1], and M.C. Lemme[1,2,*]

[1]KTH – Royal Institute of Technology, School of Information and Communication Technology, Electrum 229, 164 40 Kista, Sweden

[2]University of Siegen, Department of Electrical Engineering and Computer Science, Hölderlinstr. 3, 57076 Siegen, Germany

*corresponding author: max.lemme@uni-siegen.de, Tel: +49-271-740-4035

**Abstract**

A chip to wafer scale, CMOS compatible method of graphene device fabrication has been established, which can be integrated into the back end of the line (BEOL) of conventional semiconductor process flows. In this paper, we present experimental results of graphene field effect transistors (GFETs) which were fabricated using this wafer scalable method. The carrier mobilities in these transistors reach up to several hundred $cm^2V^{-1}s^{-1}$. Further, these devices exhibit current saturation regions similar to graphene devices fabricated using mechanical exfoliation. The overall performance of the GFETs can not yet compete with record values reported for devices based on mechanically exfoliated material. Nevertheless, this large scale approach is an important step towards reliability and variability studies as well as optimization of device aspects such as electrical contacts and dielectric interfaces with statistically relevant numbers of devices. It is also an important milestone towards introducing graphene into wafer scale process lines.

**Introduction**

In today's semiconductor technology, there is a growing trend to not only scale devices and increase their density on a wafer (more Moore), but also to increase the functionality of integrated circuits as a whole (more than Moore). Devices integrated on chip are no longer relegated to performing logic operations or memory functions, but may provide electronic and electromechanic sensing for environmental feedback or radio frequency (RF) analog data handling for high speed wireless communication. Research into graphene has progressed rapidly – and a number of potential more than Moore applications are being investigated from sensors [1], [2] to radio frequency (RF) devices [3]–[5] and photodetectors [6], [7]. Graphene can have high carrier mobility on $SiO_2$ substrates (up to 20,000 $cm^2V^{-1}s^{-1}$) [8] and shows high velocity saturation [9]. Clear current saturation and high transconductance have also been measured for large area devices fabricated using pristine (mechanically exfoliated) graphene [10]. These extraordinary electronic properties have fueled research into graphene field effect transistors (GFETs) [11], in particular for RF applications. This is due to the absence of an electronic band gap, which leads to poor on to off current ratios ($I_{on}/I_{off}$) in GFETs and limits their use in conventional digital circuits [12]. In RF devices, in contrast, switching with high $I_{on}/I_{off}$ ratios is a less stringent requirement [3], [13], [14]. At present, several experimental demonstrations of simple GFET-based circuits are available [15]–[19], as well as device models that enable further exploitation in circuit design [10], [20]–[22].

One of the most crucial steps towards commercializing graphene devices and circuits, besides demonstrating prototypes in experiments, is to establish wafer scale processes that are compatible with existing process technology. In this work, we show the feasibility of a large scale graphene process flow for fabrication of GFETs. This fabrication method can be implemented in the back end of the line (BEOL) of a CMOS compatible process, is in principal scalable to any wafer size, and could potentially allow for the integration of RF transistors and circuits or other graphene-based devices in system on chip (SoC) applications, adding system functionality.

**Experimental**

Fig. 1a shows a schematic of the GFET fabrication process, which in detail comprises of the following steps: Devices are fabricated using a silicon substrate covered with a 1.8 μm layer of thermally grown silicon dioxide ($SiO_2$) (Fig. 1a-1). Reactive ion etching (RIE) is used in order to selectively etch through the $SiO_2$ layer to form electrical contacts to the substrate (Fig. 1a-2). The RIE process uses 200 mW power at 40 mTorr. Aluminum is then deposited using e-beam evaporation into the contact holes in order to act as vias to the substrate / back gate. The aluminum via is patterned using standard photolithography and lift-off (Fig. 1a-3). RIE is similarly used to etch 640 nm deep trenches into the $SiO_2$ layer, which are then filled with an adhesion layer of 160 nm of titanium followed by 500 nm of gold to form drain and source contacts (Fig. 1a-4). Evaporation into the trenches effectively embeds the contacts in the $SiO_2$ layer, which allows the graphene to be transferred over the top of the contacts instead of forming the contacts on top of graphene. Embedding contacts before graphene transfer reduces the number of process steps after graphene is transferred, which reduces damage to the graphene layer. The drain and source contacts are patterned using lift-off. Note that the gold contacts can be replaced easily with suitable CMOS compatible metals and the lift-off process can be replaced with standard BEOL damascene technology. Graphene is then transferred to the wafer as shown in Fig. 1a-5. Commercially available chemical vapor deposited (CVD) graphene from Graphenea is used for the transfer (http://www.graphenea.com/). The graphene transfer process is as follows: first, a layer of poly(Bisphenol A) carbonate (PC) is spin coated onto the copper and graphene in a common wet transfer method [23]–[27] as shown in Fig. 2-1. Carbon residue is subsequently etched away from the back side of the copper using $O_2$ plasma. The copper/graphene/PC stack is then placed into ferric chloride in order to etch away the copper layer. This leaves a remaining graphene/PC stack floating in the etchant. The stack is then transferred to de-ionized (DI) water and 8% hydrochloric acid HCl in order to clean away remaining residues. It is then transferred to the destination wafer (Fig. 2-2). The wafer is then baked at 45˚C on a hotplate for 10 minutes in order to evaporate remaining water as well as improve the adhesion between the substrate and the graphene. Acetone is used for initial cleaning of the wafer before placing it into chloroform overnight to etch the polymer layer (Fig. 2-3). The graphene is then patterned using standard photolithography and etched using $O_2$ plasma (Fig. 1a-6). Metal evaporation is then used to deposit a 3 nm layer of aluminum onto the surface of the wafer. This layer is oxidized to act as a seed layer for subsequent deposition of 20 nm of $Al_2O_3$ using atomic layer deposition (ALD), leading to a 25 nm thick dielectric layer (Fig. 1a-7). The $Al_2O_3$ in areas of the wafer such as the contact pads are then selectively etched using photolithography and a wet chemical etchant (CD 26). A gate electrode is then deposited on top of the $Al_2O_3$ which consists of 50 of Ti and 150 nm of Au. Metal evaporation and lift-off are again used for the top gate (Fig. 1a-8). Fig. 1b shows a graphene patch transferred onto a 4" wafer. Note that the graphene transfer process is limited by the available graphene size, not by the transfer method. Finally, the chips are wire bonded and packaged for further experiments. Figure 2a shows a color enhanced scanning electron microscope (SEM) image of a wire bonded GFET device. Wire bonds are shaded orange, drain and source contacts are shaded gold, the top gate is shaded yellow, the back gate is shaded blue, and the graphene channel is shaded light blue. Fig. 2b shows a wire bonded chip inside a 24 lead dual-in-line package (DiP).

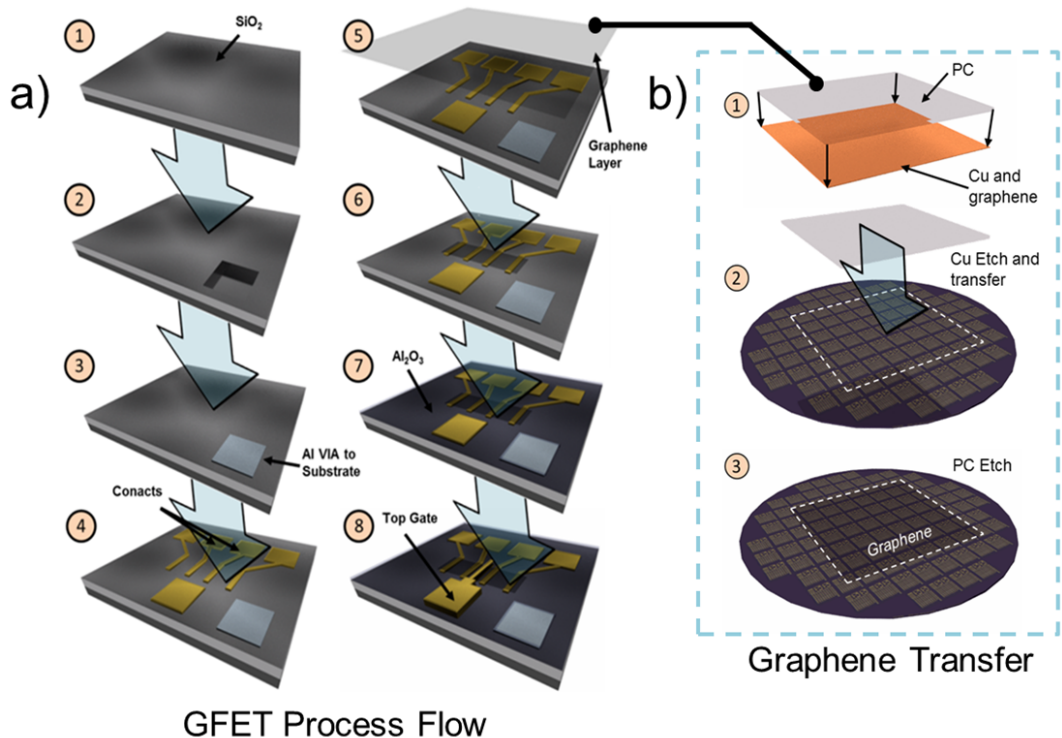

GFET Process Flow

Graphene Transfer

Figure 1: a) Schematic of the GFET process flow including the following process steps: Thermal oxidation of silicon to form SiO$_2$ (1). Via etching into the SiO$_2$ layer to the substrate using RIE (2). Filling of the via with aluminum to act as a contact pad to the substrate (3). Metal contacts patterning (4). Graphene transfer (5). Graphene patterning using O$_2$ plasma (6). Al$_2$O$_3$ gate oxide deposition on graphene (7). Top gate formation (8). b) Wafer scale transfer process. Polycarbonate (PC) spin coating on Cu foil containing graphene (1). FeCl$_3$ etching of the Cu foil and transfer of the graphene/PC stack onto the wafer (2). Etching of the PC layer in Chloroform, leaving the graphene sheet on the wafer.

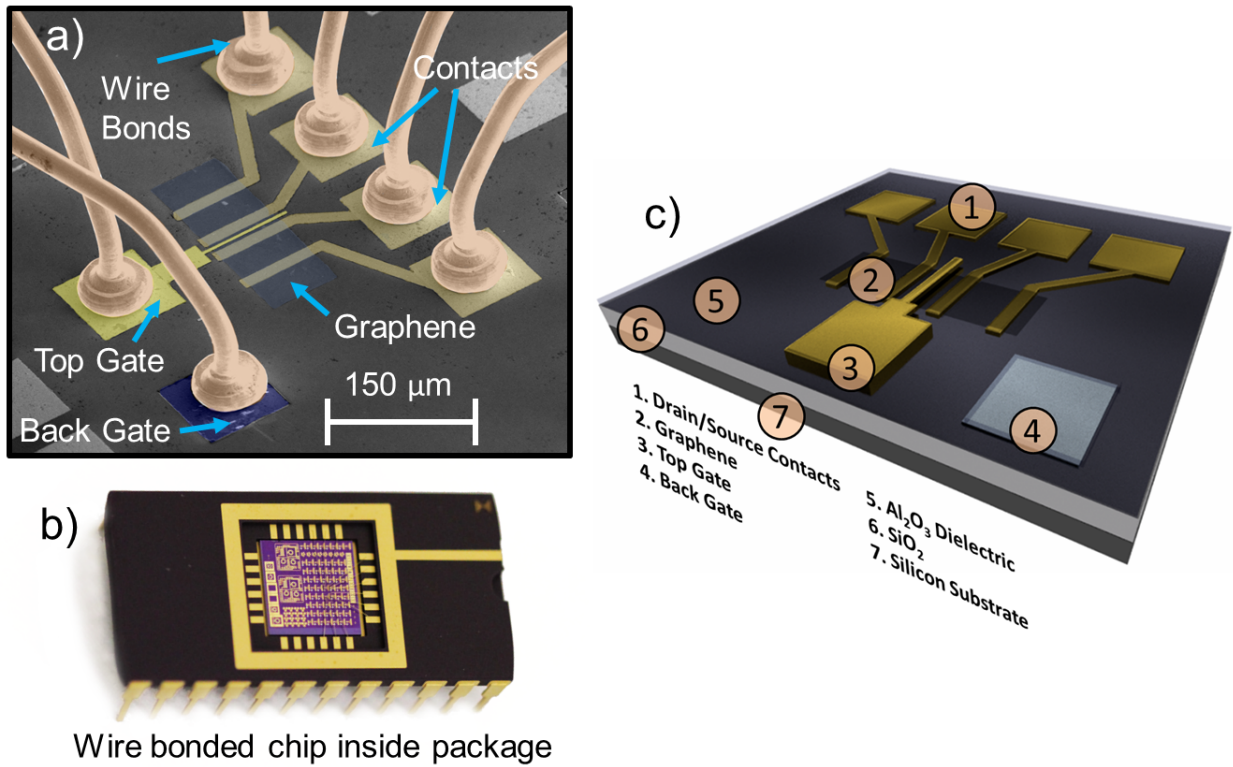

Figure 2: a) Color enhanced SEM image of a wire bonded GFET where wire bonds are shaded orange, drain and source contacts are shaded gold, the top gate is shaded yellow, the back gate is shaded blue, and the graphene channel is shaded light blue. b) Wire bonded graphene chip inside a chip package. c) Isometric representation of a GFET fabricated using the method described in Fig. 1.

**Results & Discussion**

A schematic of the fabricated GFETs is shown in Fig 2c. A bias voltage can be applied to the device using the Ti/Au drain and source contacts (1). The graphene layer forms the transistor channel between source and drain (2). The devices also provide a contact pad for the Ti/Au top gate electrode (3), which is electrically isolated from the channel by a 25 nm $Al_2O_3$ layer (5). A 1.8 µm thick $SiO_2$ layer (6) separates the graphene channel from a silicon substrate, which can be used as a global back gate (7), with an Al via providing a contact pad for electrical probing (4).

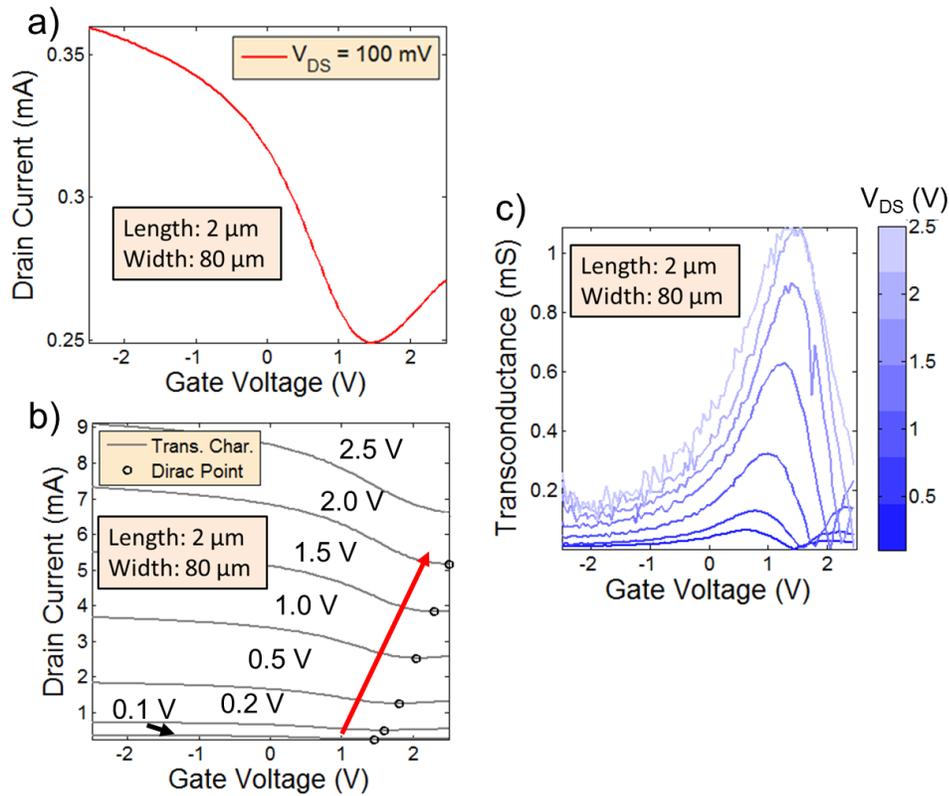

Figure 3: a) Transfer characteristics of a GFET fabricated using the large area transfer method. The device dimensions are 2 µm by 80 µm at a source-drain voltage of $V_{DS}$ = 100 mV. b) Transfer characteristics of the 2 µm length by 80 µm width GFET for $V_{DS}$ ranging from 0.1 V to 2.5 V. Black circles show the Dirac positions and the red arrow indicates the Dirac voltage shifts as $V_{DS}$ is **increased. c) Transconductance of the 2 µm length by 80 µm width GFET for $V_{DS}$ ranging from 0.1 V to 2.5 V.**

Fig. 3a shows the transfer characteristics of a GFET with a channel length of 2 µm and a channel width of 80 µm, including the characteristic 'V' shape due to graphene's ambipolar nature. The measurement in Fig. 3a was carried out with a constant drain voltage of $V_{DS}$ = 100 mV. Fig. 3b shows transfer characteristics at different drain voltages ranging from $V_{DS}$ = 0.1 V to 2.5 V for a 2 µm (length) by 80 µm (width) GFET. We observe that the point of minimum conductance (i.e. the Dirac point) shifts towards more positive gate voltages $V_G$ with more positive $V_{DS}$, as indicated by the red arrow. For long channel devices (on the order of several microns), this is expected, as the minimum current for devices with ambipolar characteristics (such as graphene) occurs when the gate modulates the Fermi level to the point where carrier injection from the source and drain equalize. Since changing $V_{DS}$ will change the carrier injection, the gate voltage at which this current minimum occurs will also change [28]. Fig. 3c shows the corresponding transconductance of the top-gate transfer characteristics for various $V_{DS}$. The observed transconductance of up to approximately 10 µS/µm is comparable to previous literature when considering the gate length, biasing conditions and carrier mobility [10]. The transconductance varies from device to device and the differences are attributed to variations carrier mobility and in contact resistances, which is one of the leading causes of poor device performance in GFETs [29].

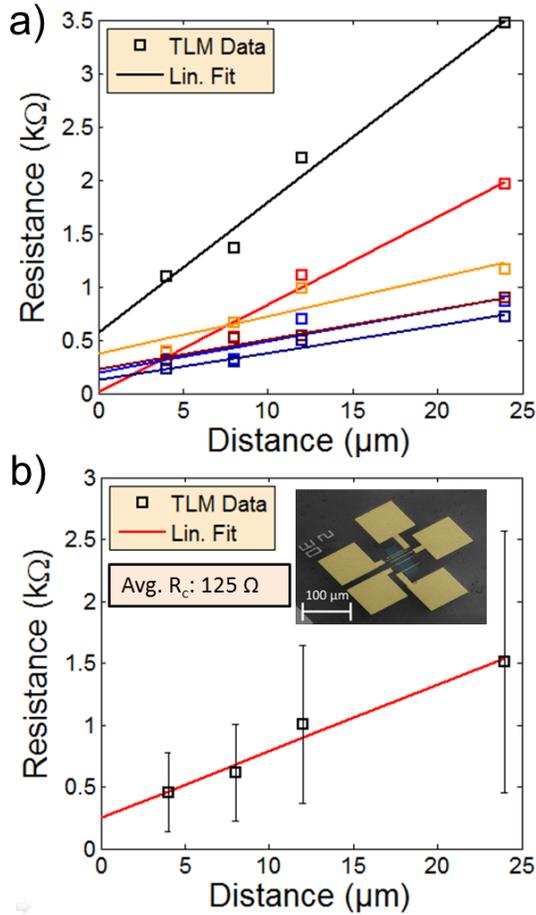

**Figure 4: a) Contact resistance extraction using several TLM structures. Squares represent resistances extracted at different channel lengths and the lines are linear fits to the resistances. b) Average resistance values plotted for different channel lengths with their corresponding standard deviation. The average extracted contact resistance is $R_C = 125\ \Omega$ per contact. The inset shows a color enhanced SEM image of a TLM structure.**

Transmission Line Model (TLM) structures were fabricated in the same process flow in order to experimentally determine the contact resistance. An example is shown in the color-enhanced SEM image in the inset of Fig. 4b, where the contacts are shaded gold and the graphene layer is shaded blue. The corresponding data is shown in Fig. 4a for 6 TLM structures. In addition, the average resistance versus channel length is shown in Fig. 4b, with error bars representing the standard deviation of resistance values from device to device. All measurements were carried out with the back gate floating. The intercept with the Y-axis yields an average contact resistance of 2·$R_C$ = 250 Ω or 125 Ω per contact with individual values from Fig. 4a ranging from approximately 7 Ω to 285 Ω per contact. Our specific contact resistance values (averaging roughly 1.5·10$^4$ Ωcm$^2$) are in line with previous literature for Ti/Au which report generally poor contact resistance values at >10$^5$ Ωcm$^2$ for devices which have not been specifically treated to reduce contact resistance [29]–[31]. The average sheet resistance for devices is 1.69 kΩ/□ which is of the same order as previously reported literature values of roughly 400-800 Ω/□ [32].

A model of the transfer characteristics of the devices was adapted from Kim et al. [31] in order to extract the charge carrier density and the carrier mobility. The model for the GFET is comprised of 3 primary equations shown in Eq.1, Eq. 2, and Eq. 3.

$$n_{tot} = \sqrt{n_0^2 + n[V_{TG} - V_{DRC}]^2} \quad \text{Eq. 1}$$

$$V_{TG} - V_{DRC} = \frac{qn}{C_{ox}} + \frac{\hbar v_F \sqrt{\pi n}}{q} \quad \text{Eq. 2}$$

$$R = R_c + \frac{N_{sq}}{n_{tot} q \mu} \quad \text{Eq. 3}$$

In these equations, $n_{tot}$ represents the charge carrier concentration, $C_{ox}$ is the capacitance of the Al$_2$O$_3$ gate dielectric, $R_c$ is the contact resistance, $v_F$ is the Fermi velocity, and $N_{sq}$ is the number of squares in the top gate region (length divided by the width). $n$ represents the portion of the charge carrier density which is dependent on the $V_{DS}$ and changes with changes in $V_{TG} - V_{DRC}$. $n_0$ is the minimum charge carrier density of the graphene defined by random dopants, impurities, and defects. Further, $\mu$ represents the mobility of the device. Extracted values from the model represent the best fit of the model to experimental data for values of $\mu$, $n_0$, and $R_c$. The maximum extracted mobility from Eq. 3 for the devices measured is 487 cm$^2$V$^{-1}$s$^{-1}$ after de-embedding the contact resistance. This value is reasonable, but not particularly high even for CVD graphene and potentially originates from several factors. One factor is impurity doping which occurs through the accumulation of residues and adsorbates during fabrication and from defects in the dielectric layer [33]. High levels of intrinsic charge carrier density are indicative of impurities [29], [31].

Another possibility is that the low mobility is a result of tears or folds in the graphene which occur during the transfer process.

Fig. 5a compares the GFET model (red line) to measured data (black dots) of a top gated GFET with L = 2 μm and W = 80 μm. The corresponding transfer characteristics and model are shown in Fig. 5b. From this model, the charge density at the location of the Dirac point was extracted to be roughly $n_{tot} = 8.1 \cdot 10^{12}$ cm$^{-2}$, which suggests significant doping and in good agreement with previous literature [10], [34]. Further, as the measurements were carried out in ambient air, part of the doping of the graphene layer is likely due to the presence of moisture, despite the presence of the top gate dielectric [33]. The model shows reasonably good agreement with the measured data, if a contact resistance of $R_C$ = 125 Ω is considered, which is consistent with the average value extracted from TLM structures (Fig. 4b).

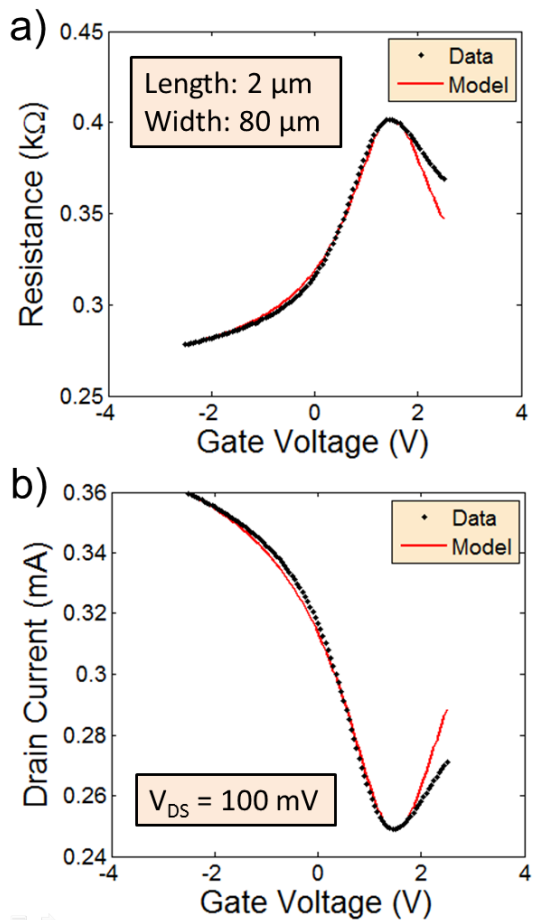

**Figure 5: a) Transfer characteristics for a 2µm length by 80 µm width device at a $V_{DS}$ of 100mV. From the model, the charge density was extracted. Black dots correspond to measured data and the red line corresponds to the model. b) Resistance versus gate voltage for the same device.**

Fig. 6a shows the output characteristics of a top gated GFET with L = 4 µm and W = 20 µm with three distinct regions (I, II, III). Fig. 6b shows the transfer characteristics of the same device with a drain bias of $V_{DS}$ = 200 mV and a Dirac point close to $V_G$ = 0 V. The red and blue dots correspond to the sections of the curve, which correspond to electron and hole charge carriers. The inset shows the leakage current of the device, which is 6 orders of magnitude lower than the drain current, and does therefore not influence the device behavior. In region I of Fig. 6a, electron conduction dominates the channel, because the gate is biased at $V_G$ = 2 V. This can be confirmed from the location of the Dirac point in Fig. 6b: the operating point is in the electron branch for positive gate voltages. The current in region I does not increase linearly with drain voltage, which can be attributed to a Schottky contact between the metal and graphene. In this device, a high contact resistance of $R_c$ ~1kΩ was extracted from the model (Eq. 3). A characteristic kink is present in region II at a drain voltage of approximately 6 V. This is the onset of the pseudo current saturation present in ambipolar long channel GFETs [10]. The kink is caused by a vanishing (or rather: minimum) carrier density at the drain, similar to the 'pinch-off' region in a semiconducting field effect transistor (FET). As $V_{DS}$ is changed, the fringing field through the underlying silicon dioxide changes the potential distribution in the channel, moving the pinch-off region (or: region of minimum conductance) along the channel. In contrast to semiconducting FETs, where the pinch off point represents the borderline between inversion charge carriers and a depleted region, it is here a border between n-type and p-type graphene, which are both conductive [20]. Thus the drain current increases again in a nearly ohmic fashion in region III after a short region of pseudo saturation.

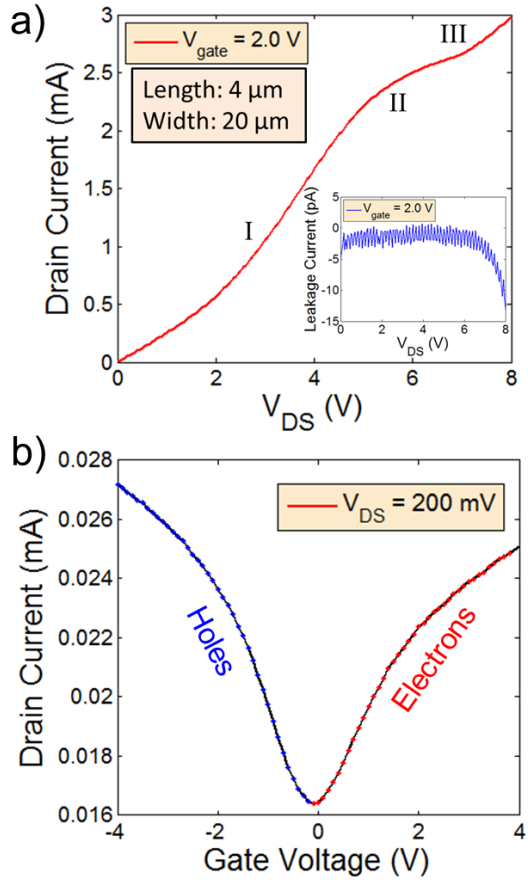

Figure 6: a) Output characteristics of GFET (L = 4 µm, W = 20 µm) at a gate voltage of $V_G$ = 2 V. The inset shows the gate leakage current of the device. b) Transfer characteristics of the same device highlighting the regions of electron and hole majority carriers.

**Conclusion**

Near wafer scale fabrication of graphene devices has been achieved using commercial CVD graphene. Electrical characterization of graphene field effect transistors shows current saturation behavior as well as transconductance values, which are within the expected range for transistors fabricated using CVD graphene. The fabrication process utilizes a scalable graphene transfer method that can potentially be integrated within the back-end-of-the-line of silicon CMOS technology. Transmission line test structures indicate the variability associated with graphene technology, a major concern for all future nanoelectronic technologies. This chip/wafer scale process enables further studies on variability and reliability of graphene devices with statistically sufficient numbers [35].

**Acknowledgements**

Support from the European Commission through a STREP project (GRADE, No. 317839), an ERC Advanced Investigator Grant (OSIRIS, No. 228229) and an ERC Starting Grant (InteGraDe, No. 307311) as well as the German Research Foundation (DFG, LE 2440/1-1 and 2-1) is gratefully acknowledged.

[20] S. A. Thiele, J. A. Schaefer, and F. Schwierz, "Modeling of graphene metal-oxide-semiconductor field-effect transistors with gapless large-area graphene channels," *J. Appl. Phys.*, vol. 107, no. 9, p. 094505, 2010.

[21] S. Fregonese, M. Magallo, C. Maneux, H. Happy, and T. Zimmer, "Scalable electrical compact modeling for graphene FET transistors," *IEEE Trans. Nanotechnol.*, vol. 12, no. 4, pp. 539–546, 2013.

[22] S. Rodriguez, S. Vaziri, A. Smith, S. Fregonese, M. Ostling, M. C. Lemme, and A. Rusu, "A Comprehensive Graphene FET Model for Circuit Design," *IEEE Trans. Electron Devices*, vol. 61, no. 4, pp. 1199–1206, 2014.

[23] L. G. De Arco, Y. Zhang, A. Kumar, and C. Zhou, "Synthesis, transfer, and devices of single-and few-layer graphene by chemical vapor deposition," *IEEE Trans. Nanotechnol.*, vol. 8, no. 2, pp. 135–138, 2009.

[24] Y. Lee, S. Bae, H. Jang, S. Jang, S.-E. Zhu, S. H. Sim, Y. I. Song, B. H. Hong, and J.-H. Ahn, "Wafer-scale synthesis and transfer of graphene films," *Nano Lett.*, vol. 10, no. 2, pp. 490–493, 2010.

[25] X. Li, Y. Zhu, W. Cai, M. Borysiak, B. Han, D. Chen, R. D. Piner, L. Colombo, and R. S. Ruoff, "Transfer of large-area graphene films for high-performance transparent conductive electrodes," *Nano Lett.*, vol. 9, no. 12, pp. 4359–4363, 2009.

[26] Y.-C. Lin, C. Jin, J.-C. Lee, S.-F. Jen, K. Suenaga, and P.-W. Chiu, "Clean transfer of graphene for isolation and suspension," *ACS Nano*, vol. 5, no. 3, pp. 2362–2368, 2011.

[27] A. Reina, H. Son, L. Jiao, B. Fan, M. S. Dresselhaus, Z. Liu, and J. Kong, "Transferring and identification of single-and few-layer graphene on arbitrary substrates," *J. Phys. Chem. C*, vol. 112, no. 46, pp. 17741–17744, 2008.

[28] S.-J. Han, Z. Chen, A. Bol, and Y. Sun, "Channel-Length-Dependent Transport Behaviors of Graphene Field-Effect Transistors," *IEEE Electron Device Lett.*, vol. 32, no. 6, pp. 812–814, Jun. 2011.